\documentclass[twocolumn,showpacs,preprintnumbers,amsmath,amssymb]{revtex4}

\usepackage{graphicx}
\usepackage{dcolumn}
\usepackage{bm}

\newcommand{\bq}{\begin{equation}}
\newcommand{\eq}{\end{equation}}
\newcommand{\bqa}{\begin{eqnarray}}
\newcommand{\eqa}{\end{eqnarray}}
\newcommand{\nn}{\nonumber \\}

\def\be     {\begin{equation}}
\def\ee     {\end{equation}}
\def\bea        {\begin{eqnarray}}
\def\eea        {\end{eqnarray}}
\def\bnn    {\begin{eqnarray*}}
\def\enn    {\end{eqnarray*}}

\begin{document}

\title{A slave-fermion gauge-theory approach of the t-J model:
Doping-induced complex magnetic structure and Z$_{2}$ spin-gapped
anomalous metal in an antiferromagnetic doped Mott insulator}

\author{ Chenglong Jia$^{*}$ and Ki-Seok Kim$^{\dagger}$ }

\affiliation{$^{*}$ Institut f\"ur Physik, Martin-Luther
Universit\"at Halle-Wittenberg, 06120 Halle (Saale), Germany }

\affiliation{$^{\dagger}$Institut de Physique Th\'eorique, CEA,
IPhT, CNRS, URA 2306, F-91191 Gif-sur-Yvette, France}

\date{\today}

\begin{abstract}
We reinvestigate a doped antiferromagnetic Mott insulator based on
the slave-fermion approach of the t-J model, where
antiferromagnetic spin fluctuations and doped holes are described
by bosonic spinons and fermionic holons, respectively. Earlier
studies have shown that an effective field theory for the doped
antiferromagnetic Mott insulator is given by a non-relativistic
fermion (holon) U(1) gauge theory for charge dynamics and a
relativistic boson (spinon) U(1) gauge theory for spin dynamics,
thus allowing an anomalous metallic phase where bosonic spinons
are gapped away from an antiferromagnetic state, analogous to the
U(1) spin liquid phase in the slave-boson approach of the t-J
model. We argue that the emergent U(1) gauge structure in this
approach is based on a simplified picture for antiferromagnetic
correlations. Considering that dynamics of doped holes frustrates
a collinear antiferromagnetic spin configuration, we show that
doped holes enhance ferromagnetic spin correlations and result in
a complex magnetic structure. The presence of such a complex spin
texture reduces the U(1) gauge structure down to Z$_{2}$, thus a
different effective field theory results for the spin-gapped
non-Fermi liquid metal at low temperatures because there are no
gapless U(1) gauge fluctuations.
\end{abstract}

\pacs{71.10.Hf, 71.10.Fd, 71.27.+a, 75.10.-b}

\maketitle

\section{Introduction}

Study on strongly correlated electrons opened a new window so
called gauge theory in modern condensed matter physics. When
electrons are weakly correlated, that is, kinetic energy is more
dominant than potential energy, a local order parameter approach
is available based on decoupling of interaction channels. This is
the heart of "classical" condensed matter physics composed of
Fermi liquid theory and Landau-Ginzburg-Wilson framework. On the
other hand, when interactions are strong enough compared with
kinetic energy, an infinite interaction limit can be a good
starting point. The presence of such a large energy scale gives
rise to a constraint in dynamics of electrons. In addition,
interesting physics now arises from the kinetic-energy
contribution in the restricted Hilbert space, and "non-local"
order parameters or more carefully, link variables instead of
on-site ones in lattice models appear as important low energy
collective degrees of freedom. These link variables can be
formulated as gauge fields, and gauge theory arises naturally for
dynamics of strongly correlated electrons.


In the present study we revisit a doped Mott insulator problem
based on one possible gauge theory approach of the slave-fermion
representation. The slave-fermion representation relies on the
fact that a spin degree of freedom of an electron is represented
by a bosonic matter, thus it is advantageous in describing
magnetic ordering of localized moments via condensation of bosonic
spinons while it has difficulty in connection with Fermi liquid.
On the other hand, the slave-boson approach resorts to the fact
that a charge degree of freedom of an electron is represented by a
bosonic matter, thus the slave-boson gauge theory has its clear
connection with the Fermi liquid theory via condensation of
bosonic holons while it does not have a "natural" description for
ordering of localized magnetic moments, compared to the
slave-fermion approach.\cite{Heavy_Fermion} An immediate question
is about statistics of spinons. At present, there is no clear
connection between the fermionic and bosonic descriptions for spin
degrees of freedom.\cite{SO5WZW}

The problem to determine which of the representations is available
seems to be associated with the nature of an undoped parent Mott
insulating phase. Although there is no obvious way to classify
undoped Mott insulating phases, it is natural to consider
symmetric and symmetry-broken Mott insulators.\cite{S_SB_MI}
Symmetric Mott insulators are usually called spin liquids, where
both fermionic\cite{PALee_Review} and bosonic\cite{Sachdev_Review}
spin descriptions are available. Symmetry-broken Mott insulators
can be discriminated further, depending on their symmetry breaking
patterns. An example of translational symmetry-broken Mott
insulators is a Mott insulator with a valence bond solid order,
and that of rotational symmetry-broken ones is an
antiferromagnetic Mott insulator. Bond-operator
formalism\cite{Sachdev_BOP} may be useful for the description of
the valence-bond-solid Mott insulator while the antiferromagnetic
Mott insulator will be described well by the Schwinger-boson
representation\cite{Book}.

In this paper we focus on doping to an antiferromagnetic Mott
insulator based on the slave-fermion approach. In early days
single-hole dynamics in the antiferromagnetically correlated spin
background was one of the main interests.\cite{SF_OLD} Hole doping
frustrates an antiferromagnetic spin configuration. Interactions
with such antiferromagnetic fluctuations give rise to a
self-energy correction of a single-hole, resulting in the fact
that it resides in four diagonal momentum points of
$(\pm\pi/2,\pm\pi/2)$. Mathematically speaking, the hopping term
in the slave-fermion representation of the t-J model, usually
called the Shraiman-Siggia term in the continuum approximation,
plays a crucial role for the single-hole dynamics. In the present
paper the role of this term is critically reinvestigated, and we
find that it plays an important role for dynamics of both holes
and spin fluctuations.

Recently, this problem was revisited by several authors, claiming
that the slave-fermion approach allows an anomalous metallic phase
with short-range antiferromagnetic correlations, where bosonic
spinons are gapped away from the antiferromagnetic Mott insulating
phase.\cite{ACL,BCS_Hubbard} An effective field theory for the
doped antiferromagnetic Mott insulator was argued to be a
non-relativistic fermion (holon) U(1) gauge theory for charge
dynamics (of gapless charged fermions around the four diagonal
points) and a relativistic boson (spinon) U(1) gauge theory for
spin dynamics (the CP$^{1}$ gauge theory of the O(3) nonlinear
$\sigma$ model for antiferromagnetic spin fluctuations), allowing
an anomalous metal phase when bosonic spinons are
gapped.\cite{NFL_SL} The presence of such an anomalous metallic
phase is attributed to
deconfinement\cite{Deconfinement_Senthil,Deconfinement_Kleinert,Deconfinement_Kim}
of compact U(1) gauge theory, where the presence of gapless
fermionic matters is expected to suppress monopole
(skyrmion\cite{Skyrmion}) excitations in two space and one time
dimensions [$(2+1)D$] although the pure compact U(1) gauge theory
does not allow such a deconfinement phase in
$(2+1)D$\cite{Polyakov}.

From a microscopic point of view, the emergent U(1) gauge
structure can be allowed when hopping of spinons and holons
between next nearest neighbor sites, i.e., hopping between same
sublattices in the square lattice has its nonzero vacuum
expectation amplitude while hopping between different sublattices
(nearest neighbor sites) vanishes in the presence of short-range
antiferromagnetic correlations (spinon singlet-pairing order). In
the present study we reexamine the spinon-holon exchange hopping
term carefully, and reveal that the emergent gauge structure is
Z$_{2}$ instead of U(1) owing to the fact that hopping of doped
holes frustrates the collinear antiferromagnetic spin
configuration and results in a complex magnetic structure. Such a
complex spin texture is reflected in nonzero expectation values of
nearest-neighbor hopping parameters of spinons and holons. In
other words, dynamics of doped holes enhances ferromagnetic spin
fluctuations, reducing the U(1) gauge symmetry, arising in the
case when only collinear antiferromagnetic correlations are
considered, down to Z$_2$.

For definiteness, we first consider the antiferromagnetic
Heisenberg model based on the Schwinger-boson representation. In
this study we allow both ferromagnetic (exchange hopping of
spinons between different sublattices) and antiferromagnetic
(singlet pairing of spinons between different sublattices)
correlations. In addition, emergence of a flux is admitted for the
ferromagnetic exchange-hopping channel. Based on the saddle-point
analysis for magnetic ordering, we find; (1) If no flux is allowed
in the exchange-hopping parameter, the vacuum expectation value of
the hopping parameter vanishes, thus our mean-field analysis
recovers the conventional Schwinger-boson mean-field theory for
collinear antiferromagnetic ordering.\cite{Book} The resulting
effective field theory is a relativistic U(1) gauge theory, i.e.,
the CP$^{1}$ gauge theory of the O(3) nonlinear $\sigma$
model.\cite{Sachdev_Read} On the other hand, (2) if $\pi$-flux is
allowed in the hopping channel, the hopping parameter between the
different sublattices has its nonzero vacuum expectation value
because the presence of $\pi$-flux enhances ferromagnetic
correlations. As a result, the U(1) gauge symmetry is down to
Z$_{2}$ since both spin-singlet pairing and exchange hopping
exist. The presence of both antiferro- and ferro- magnetic
contributions causes the magnetic ordering structure to be
complicated. We discuss physics of this possible complex magnetic
structure.

Next, we study a doped antiferromagnetic Mott insulator based on
the U(1) slave-fermion representation of the t-J model. As
discussed before, the main point is how to take the hopping-t term
into account. Performing the similar mean-field analysis as the
above, we find that the hopping parameter does not vanish even in
the uniform flux case owing to the fact that mobile doped holes
increase ferromagnetic correlations. As a result, a complex
magnetic structure such as a spiral type is expected to appear,
consistent with doping-induced spiral magnetic ordering of the
previous slave-fermion studies.\cite{SF_OLD} Enhancement of
ferromagnetic correlations induced by doped holes causes a serious
result that gapless U(1) gauge fluctuations do not exist in the
long-wave length and low-energy limits even when such bosonic
spinons become gapped to cause the fact that the spiral-type
complex magnetic structure disappears away from half filling. As a
result, this spin-gapped anomalous metallic state, where only
charge fluctuations are gapless exhibiting so called spin-charge
separation, is described by a Z$_{2}$ gauge theory instead of a
U(1) gauge theory. In this respect such a deconfinement phase is
certainly possible for $(2+1)D$, considering that even the pure
Z$_{2}$ gauge theory without matters allows its deconfinement
phase in $(2+1)D$.\cite{Fradkin_gauge_theory} We propose this
gauge-symmetry reduction in the low energy limit as one possible
mechanism for spin-charge separation in the doped
antiferromagnetic Mott insulator.

\section{Antiferromagnetic Mott insulator}

\subsection{Schwinger-boson mean-field analysis of the
antiferromagnetic Heisenberg model}

We consider the quantum antiferromagnetic Heisenberg Hamiltonian
\begin{equation}
\frac{H}{J}= \sum_{ ij } \vec{S}_{i} \cdot \vec{S}_{j} .
\end{equation}
Generally speaking, one can decompose the spin-exchange
interaction $\vec{S}_{i}\cdot \vec{S}_{j}$ with Schwinger-bosons
as follows,

\emph{(i)}  N\'{e}el scheme,
\begin{equation}
\vec{S}_{i} \cdot \vec{S}_{j} = -2\Delta_{ij}^{\dagger}\Delta_{ij}
+S^{2}
\label{eq::Neel}
\end{equation}

\emph{(ii)} Spiral scheme,
\begin{equation}
\vec{S}_{i} \cdot \vec{S}_{j} = -\Delta_{ij}^{\dagger}\Delta_{ij}
+ \chi_{ij}^{\dagger}\chi_{ij}-S/2
\label{eq::Sprial}
\end{equation}
where
$\Delta_{ij}=(b_{i\uparrow}b_{j\downarrow}-b_{i\downarrow}b_{j\uparrow})/2$
and
$\chi_{ij}^{\dagger}=(b_{i\uparrow}^{\dagger}b_{j\uparrow}+b_{i\downarrow}^{\dagger}b_{j\downarrow})/2$
are associated with antiferromagnetic and ferromagnetic
correlations, respectively.

In order to clearly show physical meaning of $\Delta_{ij}$ and
$\chi_{ij}$, it is convenient to consider classical spins. If we
choose two nearest neighbor spins
\begin{eqnarray}
b_{i}=\sqrt{2S}
         \left( \begin{array}{c}
                    1\\
                    0
                \end{array}\right),
~~~~
b_{j}=\sqrt{2S}
         \left( \begin{array}{c}
                    e^{-i\phi}\cos(\theta/2)\\
                    \sin(\theta)
                \end{array}\right) ,
\end{eqnarray}
where $(\phi,\theta)$ are spherical coordinates of spin $j$, then
\begin{eqnarray}
&& \Delta_{ij}=S\sin(\theta/2)  \\
&& \chi_{ij}=Se^{i\phi}\cos(\theta/2) \label{AB} .
\end{eqnarray}
For a ferromagnetic state ($\theta=0$), we have $\Delta=0$ and
$|\chi|=S$ while in the N\'{e}el state of $\theta=\pi$, $\Delta=S$
and $\chi=0$ result. In this respect $\Delta$ and $\chi$ represent
antiferromagnetic and ferromagnetic correlations, respectively.
Furthermore, for classical antiferromagnetic spins, when spins are
non-collinear (\emph{i.e.} $\theta \neq \pi$) where neighboring
spins have their finite overlap, $\chi$ becomes nonzero so that
$\chi$ describes canting and spiral. One may expect that
competition between antiferromagnetic $\Delta$ and ferromagnetic
$\chi$ correlations will give rise to non-collinear ordering in
the ground state of antiferromagnetically correlated spins.

Inserting Eqs. (5) and (6) into Eqs. (2) and (3),  we get
\begin{eqnarray}
& (\vec{S}_{i}\cdot \vec{S}_{j})_{N\acute{e}el} =  S^{2} \cos \theta \\
& (\vec{S}_{i}\cdot \vec{S}_{j})_{Spiral} =  S^{2} \cos \theta-S/2 .
\end{eqnarray}
It is obvious that the Spiral scheme is more favorite than the
N\'{e}el scheme  in the classical limit since it has lower ground
state energy. In the following we focus on the Spiral scheme.

\begin{figure}[b]
\begin{center}
\includegraphics[width=7cm]{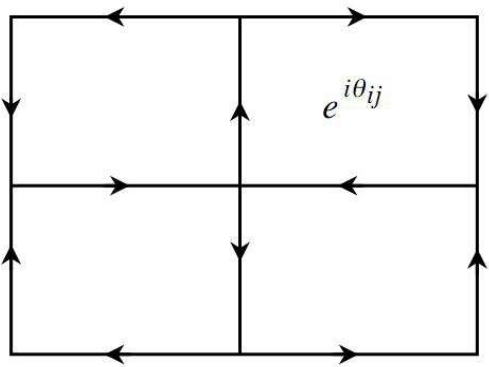}
\end{center}
\caption{Flux ansatz} \label{fig:Ek_flux}
\end{figure}

Taking into account the constraint of
$\sum_{\sigma}b_{i\sigma}^{\dagger}b_{i\sigma} = 2S$ at each site
via a Lagrange multiplier field $\lambda_{i}$, an effective
Hamiltonian in the Schwinger-boson representation of the
antiferromagnetic Heisenberg model can be written as follows
\begin{eqnarray}
\frac{H}{J} &=& -\sum_{ij,\sigma}\left( \Delta_{ij}^{*}\sigma
b_{i\sigma}b_{j\bar{\sigma}}+h.c.\right)
+\sum_{ij,\sigma}\left(\chi_{ij}b_{i\sigma}^{\dagger}b_{j\sigma}+h.c.\right) \nonumber \\
&+&
\sum_{i,\sigma}\lambda_{i}(b_{i\sigma}^{\dagger}b_{i\sigma}-2S)
+\sum_{ij}\left(|\Delta_{ij}|^{2}-|\chi_{ij}|^{2}\right) .
\label{AFH}
\end{eqnarray}
It is important to notice that the sign of the spinon-hopping term
is positive, implying that spinon-hopping contributions are
energetically unfavorable, thus its amplitude will vanish in the
uniform-flux case, as will be shown below. Physically speaking,
ferromagnetic correlations frustrating the antiferromagnetically
correlated background are prohibited in the naive mean-field
analysis. However, introduction of $\pi$-flux weakens such
frustration effects, ferromagnetic correlations contributing to
physics of antiferromagnetism.

For a mean-field analysis we replace $\Delta_{ij}$ and $\chi_{ij}$
with their saddle-point values. In the flux ansatz we parameterize
$\Delta_{ij}$ and $\chi_{ij}$ with
\begin{equation}
\Delta_{ij}= \Delta, ~~~~ \chi_{ij}= \chi e^{i\theta_{ij}}
\end{equation}
where $\theta_{ij}=\pm \theta$ if $ij$ along or against an arrow
in Fig. 1. Note that a flux is allowed only in the ferromagnetic
hopping channel. Based on this mean-field ansatz with $\lambda_{i}
= \lambda$, we obtain the following effective Hamiltonian in the
momentum space,
\begin{eqnarray}
&& \frac{H}{J} = -\frac{z\Delta}{2} \sum_{k,\sigma}\left( \sigma
\gamma_{k}b_{k\sigma}b_{\bar{k}\bar{\sigma}}+h.c.\right)
+ \lambda\sum_{k\sigma}b_{k\sigma}^{\dagger}b_{k\sigma}\nonumber \\
&&+ z\chi \sum'_{k,\sigma}\left(b_{k\sigma}^{\dagger} ~~
b_{k+Q\sigma}^{\dagger}\right) \left(\begin{array}{cc}
       \gamma_{k}\cos \theta & i \phi_{k}\sin \theta \\
       -i\phi_{k} \sin \theta & -\gamma_{k}\cos \theta
       \end{array}\right)
\left(\begin{array}{c}
       b_{k\sigma}\\
       b_{k+Q\sigma}
       \end{array}\right) \nonumber \\
&&- 2NS\lambda+\frac{z}{2}N\Delta^{2}-\frac{z}{2}N\chi^{2}
\end{eqnarray}
with $\gamma_{k}=1/z \sum_{j\in i}(e^{i k_{x}} + e^{i k_{y}})$ and
$\phi_{k}=1/z \sum_{j\in i}(e^{i k_{x}} - e^{i k_{y}})$. $z$ is a
lattice-coordination number, here $z=4$ for the square lattice.
The $'$ symbol in the momentum summation represents to perform the
summation in the folded Brillouin zone.

Performing the Bogoliubov transformation, we can diagonalize the
effective Hamiltonian as follows,
\begin{eqnarray}
\frac{H}{J} &=& \sum'_{k,s=\pm}\omega_{k}^{s}\left(
\alpha_{ks}^{\dagger}\alpha_{ks}+
\beta_{ks}^{\dagger}\beta_{ks}+1\right) \nonumber \\
&-&
2N\lambda(S+\frac{1}{2})+\frac{z}{2}N\Delta^{2}-\frac{z}{2}N\chi^{2}
,
\end{eqnarray}
where the quasiparticle spectrum is
\begin{eqnarray}
\omega_{k}^{s}= \sqrt{(\lambda
+sz\chi\Theta_{k})^{2}-(z\Delta\gamma_{k}/2)^{2}}
\end{eqnarray}
with $\Theta_{k}= \sqrt{( \gamma_{k} \cos\theta)^{2}+ (\phi_{k}
\sin\theta )^{2}}$ in the flux ansatz. Then, the resulting
mean-field free energy is given by
\begin{eqnarray}
&& F_{MF} = \frac{2}{\beta}\sum'_{ks}\ln \sinh\frac{\beta
J\omega_{k}^{s}}{2} \nn &&-2NJ\lambda(S+\frac{1}{2})
+\frac{z}{2}NJ\Delta^{2}-\frac{z}{2}NJ\chi^{2} .
\end{eqnarray}

Minimizing $F_{MF}$ with respect to $\Delta$, $\chi$, and
$\lambda$, we obtain self-consistent mean-field equations
\begin{eqnarray}
&& S+1/2 = \frac{1}{N}\sum'_{ks}\frac{\lambda+ s\chi\Theta_{k}}{\omega_{k}^{s}} (n_{k}+1/2) \\
&& \Delta = \frac{1}{N}\sum'_{ks}\frac{z\Delta\gamma_{k}^{2}}{2\omega_{k}^{s}} (n_{k}+1/2) \\
&& \chi = \frac{1}{N}\sum'_{ks}\frac{2zs\Theta_{k}(\lambda+
s\chi\Theta_{k})}{\omega_{k}^{s}} (n_{k}+1/2)   \label{MFE}
\end{eqnarray}
with $\Delta\rightarrow z\Delta/2$ and $\chi\rightarrow z\chi$.

Flux states of our interest are uniform-flux ($\theta=0$) and
$\pi$-flux ($\theta = \pi/4$) phases since they do not break
time-reversal symmetry. \emph{For the uniform phase}, our
numerical analysis shows that the lowest free energy is achieved
by $\Delta=1.1580$ and $\chi=0$. One can easily check that $\chi =
0$ is a solution for the above self-consistent equations in the
uniform-flux case since the right-hand-side of Eq. (17) is odd for
momentum $k$ if $\chi = 0$, thus it vanishes after the momentum
integration, consistent with the left-hand-side of Eq. (17). In
addition, this result is consistent with our physical picture
associated with the sign of the ferromagnetic hopping term.
Schwinger bosons are condensed to the $k^{*}=(0,0)$ state, which
corresponds to the collinear antiferromagnetic order. Staggered
magnetization is found to be $m_{0}=0.3034$, which is the same as
the Auerbach's result\cite{Book}. The ground state energy per bond
is given by
\begin{equation}
E_{0} =  - \Delta^{2} = -0.3352 \label{E0} .
\end{equation}

\emph{For the $\pi$-flux phase}, there are two branches of bosons,
$\omega_{k}^{+}$-boson and $\omega_{k}^{-}$-boson. Spinon
condensation arises in each branch when

\emph{(1) $\lambda= \Delta +\sqrt{2}\chi/2$.} At the wave vector
$k^{*}=(0,0)$ we obtain $\omega^{+}_{k^{*}}>0$ but
$\omega^{-}_{k^{*}}=0$. The self-consistent equation (\ref{MFE})
deduces $\chi=0$ corresponding to the lowest energy, which is the
same as the uniform-flux case, and $\pi$-flux doesn't give any
particular effect. N\'{e}el ordering occurs only for the
$\omega_{k}^{-}$ branch.

\begin{figure}[t]
\begin{center}
\includegraphics[width=8cm]{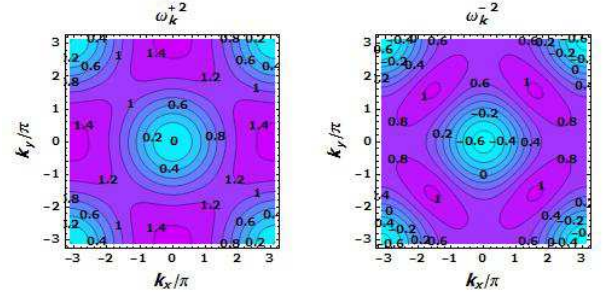}
\end{center}
\caption{ (Color online) $\omega_{k}^{+2}$ and $\omega_{k}^{-2}$
in the $\pi$-flux phase when $\lambda = \Delta -\sqrt{2}\chi/2$
with $\Delta=1.2021$ and $\chi=0.2349$, where there exists a
contour for the quasiparticle spectrum to cross from positiveness
to negativeness in $\omega_{k}^{-2}$.} \label{fig:Ek_PI}
\end{figure}

\emph{(2) $\lambda = \Delta -\sqrt{2}\chi/2$.}
$\omega_{k}^{+}$-branch bosons are condensed to the $k^{*}=(0,0)$
state, which implies that collinear antiferromagnetic ordering is
also realized within $\omega_{k}^{+}$ bosons. On the other hand,
if we examine the dispersion of low lying excitations of
$\omega_{k}^{-}$ bosons, we find $\omega_{k}^{-} \approx
\sqrt{2}\Delta/2 \left( |k|^{2}-4\sqrt{2}\chi/\Delta\right)^{1/2}$
around the zero point. That is, there is a ring divergence rather
than points provided that $\chi \neq 0$ (see Fig.
\ref{fig:Ek_PI}). Thus, for the $\omega_{k}^{-}$ branch the bosons
with $|k|^{2} \leq 4\sqrt{2}\chi/\Delta$ are all going into the
condensate portion of the boson state. In other words, the density
of $\omega_{k}^{-}$ bosons stays at $|k|^{2} \leq
4\sqrt{2}\chi/\Delta$ in momentum space, thus many kinds of
non-collinear magnetic ordering emerge. Actually, we obtain a
non-zero $\chi=0.2349$ and $\Delta=1.2021$, solving Eq.(\ref{MFE})
numerically. Moreover, comparing with the ground state energy of
the uniform-flux phase Eq.(\ref{E0}), this complex magnetic state
is found to have a lower energy indeed,
\begin{eqnarray}
E_{\pi} = -\Delta^{2}+\chi^{2}= - 0.3578 .
\end{eqnarray}
We can extract the collinear staggered magnetization from the
$\omega_{k}^{+}$ branch, and find $m_{0}=0.1043$. This value is
much smaller than that of the uniform-flux case owing to
enhancement of ferromagnetic correlations.

\subsection{Discussion}

In the previous section we have seen that ferromagnetic
correlations do not contribute to physics of collinear
antiferromagnetism in the conventional Schwinger-boson mean-field
approximation, i.e., uniform-flux ansatz. On the other hand, the
$\pi$-flux phase turns out to be more stable than the uniform-flux
phase, and the emergence of $\pi$-flux enhances ferromagnetic
correlations, causing a complex magnetic state, where there exist
various ordering wave vectors.

Nature of the magnetically ordered state in the $\pi$-flux phase
is an important question. The present semiclassical approach does
not seem to be sufficient for determining the resulting ground
state. This situation reminds us of
"order-by-disorder",\cite{Order_By_Disorder} where quantum
fluctuations split degeneracy and determine its ground state. One
possible scenario in the square lattice is that magnetic ordering
is still to show collinear antiferromagnetism despite the presence
of such "degenerate" states, i.e., many $k^{*}$s corresponding to
ordering vectors, since magnetism measured from the degenerate
states will be cancelled out after momentum integration if the
condensation amplitude at each momentum point is assumed to be the
same as each other. These contributions reduce the strength of the
collinear antiferromagnetism, as seen in the previous calculation.
This is an important effect of ferromagnetic fluctuations in the
antiferromagnetic ground state.

Existence of the complex spin texture prohibits emergence of
gapless U(1) gauge fluctuations owing to nonzero vacuum
expectation values of both spinon-hopping and spinon-pairing order
parameters. Instead, Z$_{2}$ gauge theory appears for possible
non-magnetic phases resulting from this complicated spin
pattern.\cite{Sachdev_Review} Then, deconfined spinon excitations
are expected to appear at high energies beyond their excitation
gap. Actually, dynamic spin-excitation spectra have shown that
there is large spectral weight at high energies beyond the
spin-wave approximation in undoped cuprates.\cite{AF_spinon} In
the present context this unidentified spectral weight may be
identified with deconfined spinon continuum allowed by Z$_{2}$
gauge theory.

However, it should be noted that the Z$_{2}$ gauge structure is
allowed only in the low energy limit, more precisely, $T <
T_{\chi}$, where the ferromagnetic exchange-hopping parameter has
its nonzero expectation value below $T_{\chi}$. Then, U(1) gauge
fluctuations appear above this temperature. The Z$_{2}$-U(1)
crossover makes deconfinement of spinons not clear. This suggests
to measure dynamic spin spectra carefully at high energies as a
function of temperature although it is beyond the scope of the
present mean-field study to predict this temperature accurately.

\section{Doped antiferromagnetic Mott insulator}

\subsection{A slave-fermion mean-field theory of the t-J model}

\subsubsection{A slave-fermion effective action}

We start from the t-J model for doped Mott insulators
\bqa && H = - t \sum_{ij}(c_{i\sigma}^{\dagger}c_{j\sigma} + H.c.)
+ J \sum_{ij}(\vec{S}_{i}\cdot\vec{S}_{j} - \frac{1}{4}n_{i}n_{j})
.  \nn \eqa
Based on the U(1) slave-fermion representation
\bqa && c_{i\sigma} = \psi_{i}^{\dagger}b_{i\sigma} \eqa with the
single occupancy constraint $\sum_{\sigma}
b_{i\sigma}^{\dagger}b_{i\sigma} + \psi_{i}^{\dagger}\psi_{i} =
2S$ where $S$ is a value of spin, here $S = 1/2$, one can rewrite
the spin-exchange term and electron hopping term in terms of
fermionic holon and bosonic spinon operators as follows
\bqa  && J \sum_{ij}(\vec{S}_{i}\cdot\vec{S}_{j} -
\frac{1}{4}n_{i}n_{j}) \nn && \rightarrow
\frac{J}{2}\sum_{ij}|\Delta_{ij}^{b}|^{2} -
J\sum_{ij}(\Delta_{ij}^{b\dagger}\epsilon_{\alpha\beta}b_{i\alpha}b_{j\beta}
+ H.c.)  , \nn 
&& - t \sum_{ij}(c_{i\sigma}^{\dagger}c_{j\sigma} + H.c.)
\rightarrow   t\sum_{ij}(\chi_{ji}^{\psi}\chi_{ij}^{b} + H.c.) \nn
&& - t \sum_{ij}(b_{i\sigma}^{\dagger}\chi_{ij}^{b}b_{j\sigma} +
H.c.) + t \sum_{ij}( \psi_{j}^{\dagger}\chi_{ji}^{\psi}\psi_{i} +
H.c.) , \nn \eqa where the bond operators
\bqa \Delta_{ij}^{b} = \sum_{\alpha\beta}
\epsilon_{\alpha\beta}b_{i\alpha}b_{j\beta} , ~~~ \chi_{ji}^{\psi}
= \sum_{\sigma} b_{i\sigma}^{\dagger} b_{j\sigma} , ~~~
\chi_{ij}^{b} = \psi_i \psi_j^{\dagger} \nn \eqa
describe antiferromagnetic, ferromagnetic correlations, and
hopping of doped holes, respectively. Comparing with the
Heisenberg Hamiltonian [Eq. (\ref{AFH})], one can see
$-t\chi_{ij}^{b}$ and $J\Delta_{ij}^{b}$ correspond to $J
\chi_{ij}$ and $J \Delta_{ij}$, respectively.

Using the above expressions, the t-J model in the electron
language is mapped onto exactly the same model but with
fractionalized particles such as spinons and holons in the
following way
\bqa && L_{eff} = \sum_{i}b_{i\sigma}^{\dagger}(\partial_{\tau} -
\mu)b_{i\sigma} - t
\sum_{ij}(b_{i\sigma}^{\dagger}\chi_{ij}^{b}b_{j\sigma} + H.c.)
\nn && -
J\sum_{ij}(\Delta_{ij}^{b\dagger}\epsilon_{\alpha\beta}b_{i\alpha}b_{j\beta}
+ H.c.) + \frac{J}{2}\sum_{ij}|\Delta_{ij}^{b}|^{2} \nn && +
\sum_{i}\psi_{i}^{\dagger}\partial_{\tau}\psi_{i} + t \sum_{ij}(
\psi_{j}^{\dagger}\chi_{ji}^{\psi}\psi_{i} + H.c.) +
t\sum_{ij}(\chi_{ji}^{\psi}\chi_{ij}^{b} + H.c.) \nn && + i
\sum_{i}\lambda_{i}(b_{i\sigma}^{\dagger}b_{i\sigma} +
\psi_{i}^{\dagger}\psi_{i} - 2S) , \eqa where $\lambda_{i}$ is a
Lagrange multiplier field to impose the single occupancy
constraint.

It is important to notice that ferromagnetic exchange correlations
in the spin-exchange term is not allowed in the hole-doped case.
This implies that at half-filling, this effective Lagrangian
reduces to the conventional Schwinger-boson theory instead of the
"flux" Schwinger-boson theory since the absence of holons results
in $\chi_{ij}^{b} = 0$, causing ferromagnetic correlations to
vanish, i.e., $\chi_{ij}^{\psi} = 0$. The reason why we do not
introduce the ferromagnetic correlation term from the Heisenberg
term is that when holes are doped, the presence of charge
contribution, $- \frac{1}{4} n_{i} n_{j}$, does not allow the
simple ground-state-energy analysis even for the case of classical
spins, performed in the study of the Heisenberg model, thus it is
not obvious that introduction of such contributions lowers the
ground state energy. Here, we focus on the hopping term of the t-J
model instead, giving rise to ferromagnetic spin correlations
induced by doped holes. This doping-induced
ferromagnetic-correlation term will play an important role in the
structure of magnetic ordering. As seen from its sign, opposite to
the half-filled case, it will contribute even in the uniform-flux
case, and modify the collinear antiferromagnetism drastically.

In the following we will consider the $\pi$-flux phase although we
comment on the uniform-flux case briefly. There are two reasons
for considering such a flux phase. The $\pi$-flux phase is
time-reversal symmetry-preserving, thus physically allowed. More
importantly, the $\pi$-flux ansatz gives rise to four hole
pockets, consistent with the previous studies\cite{SF_OLD}, where
the spinon-holon exchange hopping term (Shraiman-Siggia term) is
taken into account in the context of the perturbation theory,
revealing that holons reside in the four diagonal points. Although
we do not know the connection between our flux-phase result and
the previous perturbation theories, the $\pi$-flux ansatz
reproduces such hole pockets.\cite{Luttinger_Theorem} On the other
hand, the uniform-flux phase does not allow the hole pockets.

\subsubsection{$\pi$-flux ansatz}

\begin{figure}[t]
\begin{center}
\includegraphics[width=8cm]{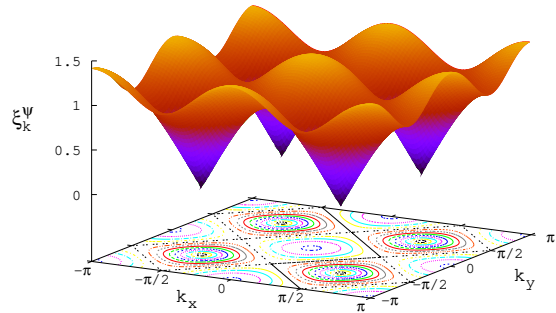}
\end{center}
\caption{ (Color online) Energy dispersion of fermionic holes,
where $Min[\xi_{k}^{\psi}]=0$ results at $k=(\pm \pi/2, \pm
\pi/2)$.} \label{fig:Ek}
\end{figure}

Four order parameters are assumed to be as follows for the
mean-field analysis
\bqa && \Delta_{ij}^b \rightarrow \Delta_b , ~~~~~
 i \lambda_{i} \rightarrow \lambda , \nn &&
\chi_{ij}^{b} \rightarrow \chi_{b}e^{-i\phi_{ij}^{b}} , ~~~~~
 \chi_{ij}^{\psi} \rightarrow \chi_{\psi}
e^{-i\phi_{ij}^{\psi}} , \eqa
decoupling the slave-fermion effective Lagrangian, where the
spin-singlet pairing order parameter and Lagrange multiplier field
are assumed to be uniform and real while the hopping parameters of
spinons and holons have their phase contributions. A nonzero value
of $\Delta_b$ corresponds to short-range antiferromagnetic
correlation as discussed before, and a nonzero $\chi_b$
($\chi_\psi$) implies hole (spinon) mobility or ferromagnetic
correlation, which deforms the spin order in the presence of hole
doping. Together with the $\pi$-flux ansatz
\bqa && \chi_{ij}^{b}\chi_{jk}^{b}\chi_{kl}^{b}\chi_{li}^{b} =
\chi_{b}^{4}e^{- i\sum_{\square}\phi_{ij}^{b}} , \nn &&
\chi_{ij}^{\psi}\chi_{jk}^{\psi}\chi_{kl}^{\psi}\chi_{li}^{\psi} =
\chi_{\psi}^{4}e^{- i\sum_{\square}\phi_{ij}^{\psi}} , \nn &&
\sum_{\square}\phi_{ij}^{b} = \sum_{\square}\phi_{ij}^{\psi} = \pi
\eqa
the free energy can be decomposed into the charge (fermion) and
spin (boson) sectors,

\emph{(1) Fermion sector}

\bqa && F_{\psi} = - \frac{1}{\beta}\sum_{ks}^{'} \ln \Bigl\{2
\cosh\Bigl( \frac{\beta E_{ks}^{\psi}}{2}\Bigr) \Bigr\} \nn && +
2t\chi_{\psi}\sum'_{k}\sqrt{(\gamma_{k}\cos\Theta)^{2} +
(\varphi_{k}\sin\Theta)^{2}} \nn && = -
\frac{1}{\beta}\sum_{k}^{'} \Bigl[ \ln \Bigl\{2 \cosh\Bigl( \frac{
\beta }{2} [2t\chi_{\psi}\sqrt{(\gamma_{k}\cos\Theta)^{2} +
(\varphi_{k}\sin\Theta)^{2}} \nn && + \lambda] \Bigr) \Bigr\} +
\ln \Bigl\{2 \cosh\Bigl( \frac{\beta}{2} [-
2t\chi_{\psi}\sqrt{(\gamma_{k}\cos\Theta)^{2} +
(\varphi_{k}\sin\Theta)^{2}} \nn && + \lambda] \Bigr) \Bigr\}
\Bigr] + 2t\chi_{\psi}\sum'_{k}\sqrt{(\gamma_{k}\cos\Theta)^{2} +
(\varphi_{k}\sin\Theta)^{2}} \eqa

\emph{(2) Boson sector}

\bqa && F_{B} = \frac{1}{\beta}\sum_{ksl}^{'} \ln
\Bigl\{2\sinh\Bigl(\frac{\beta E_{ksl}^{b}}{2}\Bigr)\Bigr\} \nn &&
= \frac{2}{\beta}\sum_{k}^{'} \Bigl[ \ln
\Bigl\{2\sinh\Bigl(\frac{\beta}{2}
\bigl\{\bigl(2t\chi_{b}\sqrt{(\gamma_{k}\cos\Theta)^{2} +
(\varphi_{k}\sin\Theta)^{2}} \nn && - \mu + \lambda \bigr)^{2} -
\bigl(4J \Delta_{b}\gamma_{k}\bigr)^{2}\bigr\}^{1/2} \Bigr)\Bigr\}
\nn && + \ln \Bigl\{2\sinh\Bigl(\frac{\beta}{2}
\bigl\{\bigl(-2t\chi_{b}\sqrt{(\gamma_{k}\cos\Theta)^{2} +
(\varphi_{k}\sin\Theta)^{2}} \nn && - \mu + \lambda\bigr)^{2} -
\bigl(4J \Delta_{b}\gamma_{k}\bigr)^{2}\bigr\}^{1/2} \Bigr)\Bigr\}
\Bigr] , \eqa
where
\bqa && \gamma_{k} = \cos k_{x} + \cos k_{y} , ~~~ \varphi_{k} =
\cos k_{x} - \cos k_{y} , ~~~ 4\Theta = \pi . \nn \eqa
Noted from the energy dispersion $\xi_{k}^{\psi} =
\sqrt{(\gamma_{k}\cos\Theta)^{2} + (\varphi_{k}\sin\Theta)^{2}}$
of fermions (Fig. \ref{fig:Ek}), one can see holes are pocketed
around the diagonal points, ($\pm \pi/2, \pm \pi/2$).


\subsubsection{Self-consistent equations}

In this study we focus on the properties at zero temperature.
Minimizing the free energy with respect to $\Delta_b$, $\chi_b$,
$\chi_{\psi}$, $\lambda$ and $\mu$, we obtain the following
self-consistent equations,

\emph{(1) Fermion sector}
\bqa
 && \delta =  \frac{1}{2N_{L}} \sum_{\xi_{k}^{\psi} \leq - \mu/2t\chi_{\psi}} 1 \label{hole} \\
&& \chi_b =-\frac{1}{4N_{L}} \sum_{\xi_{k}^{\psi}\leq - \mu /2t\chi_{\psi}} \xi_{k}^{\psi} \label{chib}
\eqa
Above equations clearly show that holons are most densely located
within the Dirac cones at $k=( \pm \pi/2, \pm \pi/2)$. For small
doping we have $\delta = \frac{1}{2\pi} \left(
\frac{\mu}{2t\chi_{\psi}} \right)$ and $\chi_b = -
\frac{\sqrt{2\pi}}{3} \delta^{3/2}$, as derived in appendix.

\emph{(2) Boson sector}

\bqa &&  \frac{1}{2 N_{L}}\sum'_{k,s=\pm}
\frac{-s\xi_k^{\psi} \bigl(\lambda + 2 s t\chi_{b}\xi_k^{\psi} )}{E^b_s(k)}= \chi_{\psi}  \\
&& \frac{1}{N_{L}}\sum'_{k,s=\pm} \frac{8J\Delta_b
\gamma_{k}^{2}}{E^b_s(k)}  = \Delta_b
\\
&& \frac{1}{N_{L}}\sum'_{k,s=\pm}
\frac{\bigl(2st\chi_{b}\xi_k^{\psi} + \lambda
\bigr)}{E^b_s(k)} = (2S + 1) - \delta
\eqa
with the spinon-quasiparticle spectrum $E^b_s (k) = \sqrt{\bigl(2
s t\chi_{b}\xi_k^{\psi} + \lambda \bigr)^{2} - \bigl(4J
\Delta_{b}\gamma_{k}\bigr)^{2}}$. It is clear that these
saddle-point equations are reduced to the conventional
Schwinger-boson mean-field equations in the zero-doping limit
$(\delta \rightarrow 0)$ since $\chi_{b}$ vanishes from Eq. (31)
and $\chi_{\psi}$ disappears accordingly from the summation for $s
= \pm$ in Eq. (32). Hole doping effects reduce the amplitude of an
effective spin from $2S$ to $2S - \delta$, as shown in Eq. (34),
and increase ferromagnetic correlations, as shown in Eq. (32).
Accordingly, antiferromagnetic correlations become weaken, seen
from Eq. (33).

In the mean-field treatment magnetic ordering is described by
condensation of Schwinger bosons. Here, we introduce
$\chi_{BC}^{\psi}$, $\Delta_{BC}^b$ and $n_{BC}$ to describe the
contribution from the $E^b_s(k^*)=0$ states, which are taken into
account separately:

\bqa &&  \frac{1}{2 N_{L}}\sum'_{k,s=\pm} \frac{-s\xi_k^{\psi}
\bigl(\lambda + 2 s t\chi_{b}\xi_k^{\psi} )}{E^b_s(k)} \nn && +
\chi^{\psi}_{BC}\sum_{k^*,s=\pm}\frac{-s\xi_{k^*}^{\psi}
\bigl(\lambda
+ 2 s t\chi_{b}\xi_{k^*}^{\psi} )}{\lambda}= \chi_{\psi}  \\
&& \frac{1}{N_{L}}\sum'_{k,s=\pm} \frac{8J\Delta_b
\gamma_{k}^{2}}{E^b_s(k)} + \Delta^b_{BC}\sum_{k^*,s=\pm} \frac{8J\Delta_b
\gamma_{k^*}^{2}}{\lambda} = \Delta_b
\\
&& \frac{1}{N_{L}}\sum'_{k,s=\pm}
\frac{\bigl(2st\chi_{b}\xi_k^{\psi} + \lambda \bigr)}{E^b_s(k)} +
n_{BC} \sum_{k^*,s=\pm}\frac{\bigl( 2st\chi_{b}\xi_{k^*}^{\psi} +
\lambda \bigr )}{\lambda} \nn && = (2S + 1) - \delta . \label{s}
\eqa
If $E^b_s(k^*)=0$ and $\sum_{k^*,s=\pm}n_{BC}(
2st\chi_{b}\xi_{k^*}^{\psi} + \lambda \bigr )/\lambda$ is finite,
spinons are condensed to the $E^b_s(k^*)=0$ states, and the system
possesses a magnetic long-range order. The ordering wave vector
$k^*$ is determined by $E^b_s(k^*)=0$, and especially, $k^*=(0,0)$
corresponds to long-range anti-ferromagnetic ordering. For
definiteness, we have checked that our treatment of these
mean-field equations recovers the uniform-flux Schwinger-boson
mean-field theory correctly. At half-filling the spinon spectrum
of $E^b_{+}(k) = E^b_{-}(k)$ is gapless at $k=(0,0)$ which ensures
their Bose condensation. Solving Eq. (\ref{s}), we see that the
spontaneous staggered magnetization is given by $\langle S_i^z
\rangle = n_{BC} \approx 0.3$, recovering the half-filled result
completely.


\subsubsection{Phase diagram}


Now, let's consider how the antiferromagnetic long-range order
evolves with doping. Due to the $\pi$-flux order, there are two
types of bosons, $B_{+}$ with energy $-E^b_{+}(k)$ and $B_{-}$
with energy $-E^b_{-}(k)$. Away from half-filling holes are
constrained within the Dirac cones, thus Eq. (\ref{chib}) gives
$\chi_b \leq 0$, which deduces $B^b_{+}$ has lower energy than
$B^b_{-}$ at finite doping, i.e., $-E^b_{+}(k)  < -E^b_{-}(k)$.
When $\lambda=8J\Delta_b + 2 \sqrt{2} t \chi_b$ is satisfied, we
have $E^b_{-}(0)=0$ for $B_{-}$ bosons, thus their Bose
condensation results at the wave vector $k^*=(0,0)$, implying an
antiferromagnetic order. For $B_{+}$ bosons within the same value
of $\lambda$, one can easily check that $1/E^b_{+}(k^*)$ shows
divergence along a ring $k^* \in \{ k | E^b_{+}(k) =0\}$. In other
words, the density of $B_{+}$ spinons resides at large number of
$k^*$ states, thus many kinds of incommensurate magnetic ordering
emerge and such a complex spin configuration deviates from the
collinear antiferromagnetic ordering, similar to the half-filling
case in the mean-field approximation.

\begin{figure}[t]
\begin{center}
\includegraphics[width=7cm]{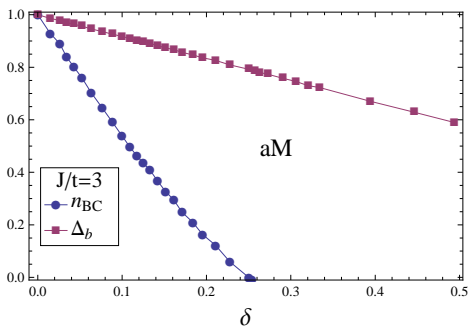}
\end{center}
\caption{ (Color online) Doping dependence of $\Delta_b$ and
$n_{BC}$ with $t/J = 3$.\cite{Spin_Value} In order to clearly show
the tendency, $\Delta_b$ and $n_{BC}$ are scaled with the values
at half-filling. The spin-gapped anomalous metallic phase is
marked as aM.} \label{fig:Phase}
\end{figure}

With increasing doping and $t$, electrons become more active,
implying that ferromagnetic correlations described by $\chi_b$ and
$\chi_{\psi}$ should also increase. However, the antiferromagnetic
order represented by the condensation amplitude $n_{BC}$ becomes
more suppressed with large $t/J$, which is confirmed by the
numerical results. Furthermore, by numerically solving the
self-consistent equations for given value of $t/J=3$,
Fig.\ref{fig:Phase} exhibits that antiferromagnetic ordering
disappears completely above $\delta_c \approx 0.25$\cite{Spin_Value}
while the spin-singlet order persists up to much larger hole
concentration  ($\Delta_b \neq 0$) away from the antiferromagnetic
order. We identify this intermediate non-magnetic spin-gapped phase
with an anomalous metal, as will be discussed in the next section.

Compared to the uniform-flux case although not shown in the
present paper, we have qualitatively the same phase diagram. This
originates from the fact that the sign of the ferromagnetic
correlation term is negative opposite to the Heisenberg model,
thus such contributions are energetically favorable even in the
uniform-flux phase. However, the presence of the $\pi$-flux order
enhances ferromagnetic correlations as seen in the study of the
Heisenberg model. As a result, the collinear antiferromagnetic
order is killed more rapidly via hole doping in the $\pi$-flux
phase. We have also examined the role of frustrated hopping
effects, i.e., next-nearest-neighbor hopping terms. Such
frustrated hopping effects modify bare dispersions for spinons and
holons. However, we do not find any qualitative changes in the
phase diagram since the presence of $\pi$-flux dominates over the
band-modification effect.

\subsection{Spin-gapped anomalous metal}

As shown in the phase diagram of Fig. 4, hole doping kills the
antiferromagnetic order and results in a paramagnetic state, where
bosonic spinons are gapped but short-range antiferromagnetic
correlations still remain. We identify this phase with an
anomalous spin-gapped metal since charge excitations carried by
holons are gapless, but spin fluctuations are gapped, thus
exhibiting spin-charge separation.

To understand physics of this anomalous metallic phase, it is
necessary to derive an effective field theory. Considering holons
first, one can expand the holon field as
\bqa  \Psi(x) & = & e^{ik_{1}x}\psi_{1}(x) +
e^{-ik_{1}x}\psi_{\bar{1}}(x) \nn & + & e^{ik_{2}x}\psi_{2}(x) +
e^{-ik_{2}x}\psi_{\bar{2}}(x)   \eqa
around the diagonal points $k_{1} = (\pi/2,\pi/2)$ and $k_{2} =
(\pi/2,-\pi/2)$, where the holon spectrum is linearized as
\bqa && E_{k}^{\psi} =
2t\chi_{\psi}\sqrt{\cos^2 k_x + \cos^2 k_y }  \approx
v_{\psi}\sqrt{k_{x}^{2} + k_{y}^{2}}   \eqa
with its velocity $v_{\psi} = 2t\chi_{\psi}$. The holon spinor is
given by
\bqa && \psi_{1}(x) = \left(\begin{array}{c} \psi_{1e}(x) \\
\psi_{1o}(x) \end{array}
\right) , ~~~~~ \psi_{2}(x) = \left(\begin{array}{c} \psi_{2o}(x) \\
\psi_{2e}(x) \end{array} \right) ,   \eqa
where $e$ and $o$ represent even and odd sites, respectively. Such
two-component spinors can be combined to form a single
four-component spinor as
\bqa && \psi(x) = \left(\begin{array}{c} \psi_{1}(x) \\
\psi_{2}(x) \end{array}
\right) = \left(\begin{array}{c} \psi_{1e}(x) \\ \psi_{1o}(x) \\
\psi_{2o}(x) \\ \psi_{2e}(x) \end{array} \right) . \eqa

An important point is that U(1) gauge fluctuations become gapped
due to condensation of a "charge" $2$ scalar field, here the
ferromagnetic hopping parameter, reduced to gapped Z$_{2}$ gauge
fields, as discussed previously. In other words, hole doping
causes magnetic frustration effects, giving rise to ferromagnetic
correlations. This reduces the emergent U(1) gauge structure down
to Z$_{2}$. As a result, we find the following effective field
theory for the anomalous Z$_{2}$ spin-gapped metal phase
\bqa && {\cal L}_{\psi} = \bar{\psi} \gamma_{\mu}(\partial_{\mu} -
iA_{\mu})\psi - \mu_{r} \bar{\psi}\gamma_{0}\psi , \eqa
where Dirac gamma matrices satisfying the Clifford algebra
$[\gamma_{\mu},\gamma_{\nu}]_{+} = 2\delta_{\mu\nu}$ are given
by\cite{Don_Kim}
\bqa && \gamma_{0} = \left(\begin{array}{cc} \sigma_{3} & 0 \\
0 & - \sigma_{3} \end{array} \right) , ~ \gamma_{1} = \left(\begin{array}{cc} \sigma_{2} & 0 \\
0 & - \sigma_{2} \end{array} \right) , ~ \gamma_{2} = \left(\begin{array}{cc} \sigma_{1} & 0 \\
0 & - \sigma_{1} \end{array} \right)  , \nonumber \eqa
and $\mu_{r}$ is a renormalized chemical potential associated with
hole doping. $A_{\mu}$ is an electromagnetic field. Note that
gapped spinon excitations are ignored for low energy physics of
this metallic phase. Then, Eq. (42) is our effective field theory
for the anomalous metallic state.

An immediate question is about the transport property in this
metallic phase. Since there is no scattering mechanism with
gapless gauge fluctuations in the above effective field theory,
Fermi liquid behavior is expected to appear in the low temperature
limit, i.e., $\delta \rho(T) = \rho(T) - \rho(T=0) \sim T^{2}$
with resistivity $\rho(T)$. Although this is true in the present
description actually, such a Fermi liquid behavior in transport is
not a full story.

As discussed in the Heisenberg model, there is an energy scale
$T_{\chi}$ associated with the crossover behavior from Z$_{2}$ to
U(1) since ferromagnetic contributions are nonzero below this
temperature. Although we cannot determine this temperature
satisfactorily in our mean-field analysis, gapless U(1) gauge
fluctuations should arise above this crossover temperature. Then,
scattering with such gapless gauge excitations would lead to
non-Fermi liquid physics in this regime. Frankly speaking, there
are various regimes at finite temperatures, associated with not
only the Z$_{2}$-U(1) crossover temperature but also so called the
spin-gap temperature $T_{b}$, where bosonic spinons are
effectively critical above this temperature. $T_{b}$ should not be
confused with $T_{\Delta}$, where spinon-singlet pairing order
appears below this temperature, much larger than the spin-gap
temperature $T_{b}$. In this respect we should consider both
energy scales, and classify various finite temperature regimes. In
the present paper we discuss only low energy physics of this
metallic phase.

\section{Discussion and summary}

\subsection{Coherent electron quasiparticle excitations near the
diagonal points}

The present slave-fermion mean-field description cannot explain
the emergence of sharply defined electron quasiparticles near the
diagonal points and their Fermi liquid behavior because electrons
decay into spinons and holons in the Z$_{2}$ anomalous metal
phase.\cite{ARPES} This is the main criticism in this kind of
decomposition approach if one tries to connect the spin-charge
separation scenario with pseudogap physics of high T$_c$ cuprates.
However, if we consider the regime of $T > T_{\chi}$, the
situation is not perfectly clear owing to the presence of gapless
gauge fluctuations beyond the mean-field description since such
gauge fluctuations cause attractive interactions between holons
and spinons, binding them, which may allow a resonance state
corresponding to an electron-like excitation although it differs
from confinement. Actually, this was a little bit demonstrated,
where noncompact gauge fluctuations give rise to attractive
interactions between "particle" and "hole" (holon and spinon in
the electron spectrum), increasing coherence.\cite{Nodal_electron}
We should confess that this argument is far from consensus even in
the qualitative point of view.

In the low temperature regime, more precisely in the zero
temperature limit where we have studied in this paper, the
situation becomes worse because even gapless U(1) gauge
fluctuations do not exist owing to the gauge-symmetry reduction.
However, we argue that such an electron excitation may still
appear as a resonance state between a spinon and holon. To see
this resonance state, we revisit the hopping-t term in the
slave-fermion representation of the t-J model. We rewrite the
hopping term as follows
\bqa && - t
\sum_{ij}(b_{i\sigma}^{\dagger}f_{i}f_{j}^{\dagger}b_{j\sigma} +
H.c.) \rightarrow t \sum_{ij}(c_{i\sigma}^{\dagger}c_{j\sigma} +
H.c.) \nn && - t \sum_{ij}(c_{i\sigma}^{\dagger}
f_{j}^{\dagger}b_{j\sigma} + H.c.) - t
\sum_{ij}(b_{i\sigma}^{\dagger}f_{i}c_{j\sigma} + H.c.) \nn && +
t\sum_{ij}(\chi_{ji}^{\psi}\chi_{ij}^{b} + H.c.) - t
\sum_{ij}(b_{i\sigma}^{\dagger}\chi_{ij}^{b}b_{j\sigma} + H.c.)
\nn && + t \sum_{ij}( \psi_{j}^{\dagger}\chi_{ji}^{\psi}\psi_{i} +
H.c.) . \eqa
It should be noted that the electron operator $c_{i\sigma}$ is
different from the original one since it is a Hubbard-Stratonovich
field although it is a fermion. Rather, it should be understood as
a collective fermionic excitation in the spinon-holon liquid. We
identify this fermion field as an electron, which is expected to
be seen in the low energy regime. Note that hopping of such low
energy electrons is frustrated, viewed from its sign.

The electron-spinon-holon exchange term reminds us of the
slave-fermion representation of the Anderson lattice model, where
localized spins are represented in slave-fermions.\cite{SF_ALM} To
find a self-energy correction from this "hybridization" coupling
seems to be an interesting and important problem since the
spin-charge separation scenario may explain the coherent nodal
electron excitations.

\subsection{d-wave superconductivity}

Another interesting subject is how to describe d-wave
superconductivity in this context. Considering the slave-boson
case, one may regard this study as a straightforward job. In the
slave-boson theory condensation of bosonic holons results in
superconductivity in the presence of spinon singlet pairing since
they form Cooper pairs. However, this structure cannot be applied
to the slave-fermion context. Fist of all, charged holons are
fermionic, thus their condensation is forbidden. Then, the next
choice will be pairing of such fermionic holons. Actually, such
pairing interactions are usually attributed to the presence of
gapless U(1) gauge fluctuations since they cause attractive
interactions between holons living in different sublattices. What
we would like to do is to construct a mean-field theory for such
pairing order from the microscopic model itself. However, if we
start from the t-J model, such an attractive pairing interaction
channel is difficult to find since the exchange term is
represented only in the bosonic-spinon language owing to the
single occupancy constraint, analogous to the fact that the J term
is also expressed in terms of only fermionic spinons in the
slave-boson context.

To avoid this difficulty, we start from so called BCS-Hubbard
model\cite{BCS_Hubbard_SR}
\bqa && H = - t \sum_{ij}(c_{i\sigma}^{\dagger}c_{j\sigma} + H.c.)
-
\sum_{ij}(\Delta_{ij}^{\dagger}\epsilon_{\alpha\beta}c_{i\alpha}c_{j\beta}
+ H.c.) \nn && + \frac{1}{g}\sum_{ij}|\Delta_{ij}|^{2} + U
\sum_{i}n_{i\uparrow}n_{i\downarrow} \eqa
where the Hubbard-Stratonovich transformation for the Cooper
channel is performed before the constraint is applied. This model
has its own interest in the fact that we can investigate what
happens in BCS superconductivity, increasing local interactions
$U$. Our strategy is that such strong correlation effects are
introduced using the slave-fermion representation in the large-$U$
limit. Then, the pairing term is expressed in the slave-fermion
representation as
\bqa && -
\sum_{ij}(\Delta_{ij}^{\dagger}\epsilon_{\alpha\beta}c_{i\alpha}c_{j\beta}
+ H.c.) \nn && = -
\sum_{ij}(\Delta_{ij}^{\dagger}\epsilon_{\alpha\beta}b_{i\alpha}b_{j\beta}
\psi_{i}^{\dagger}\psi_{j}^{\dagger} + H.c.) \nn && \rightarrow
\sum_{ij}(\Delta_{ij}^{\dagger}\Delta_{ij}^{\psi}\Delta_{ij}^{b\dagger}
+ H.c.)  - \sum_{ij}(\Delta_{ij}^{\dagger} \Delta_{ij}^{b\dagger}
\epsilon_{\alpha\beta}b_{i\alpha}b_{j\beta} + H.c.) \nn && -
\sum_{ij} (\Delta_{ij}^{\dagger}\Delta_{ij}^{\psi}
\psi_{i}^{\dagger}\psi_{j}^{\dagger} + H.c.) ,  \eqa
where pairing interactions are renormalized due to Hubbard
interactions. Replacing $\Delta_{ij}^{\dagger}
\Delta_{ij}^{b\dagger}$ and
$\Delta_{ij}^{\dagger}\Delta_{ij}^{\psi}$ with $\Delta_{ij}^{b}$
and $\Delta_{ij}^{\psi}$, respectively, we find the slave-fermion
effective Lagrangian for the BCS-Hubbard model in the large-$U$
limit
\bqa && L_{eff} = \sum_{i}b_{i\sigma}^{\dagger}(\partial_{\tau} -
\mu)b_{i\sigma} - t
\sum_{ij}(b_{i\sigma}^{\dagger}\chi_{ij}^{b}b_{j\sigma} + H.c.)
\nn && - \sum_{ij}(\Delta_{ij}^{b\dagger}
\epsilon_{\alpha\beta}b_{i\alpha}b_{j\beta} + H.c.) \nn && +
\sum_{i}\psi_{i}^{\dagger}\partial_{\tau}\psi_{i} + t \sum_{ij}(
\psi_{j}^{\dagger}\chi_{ji}^{\psi}\psi_{i} + H.c.) \nn && -
\sum_{ij} (\Delta_{ij}^{\psi} \psi_{i}^{\dagger}\psi_{j}^{\dagger}
+ H.c.) \nn && +
\sum_{ij}(\frac{\Delta_{ij}^{\psi\dagger}\Delta_{ij}^{b\dagger}}{\Delta_{ij}}
+ H.c. + \frac{1}{g} |\Delta_{ij}|^{2}) +
t\sum_{ij}(\chi_{ji}^{\psi}\chi_{ij}^{b} + H.c.) \nn && + i
\sum_{i}\lambda_{i}(b_{i\sigma}^{\dagger}b_{i\sigma} +
\psi_{i}^{\dagger}\psi_{i} - 1) . \eqa

Although investigation of this effective Lagrangian is beyond the
scope of this paper, one can read an important aspect of this
Lagrangian immediately. An observed dome-shaped superconductivity
line is expected to appear since $\Delta_{ij}^{b}$ vanishes in the
zero doping limit owing the absence of doped holes, causing
$\Delta_{ij}^{\psi} = 0$ while $\Delta_{ij}^{\psi}$ disappears at
large doping, killing $\Delta_{ij}^{b}$.

\subsection{Summary}

In this paper we have considered dynamics of doped holes in the
antiferromagnetically correlated spin background. The point is
that doped holes frustrate the collinear antiferromagnetic spin
configuration, resulting in more complex spiral-like spin
patterns. This reduces the U(1) gauge symmetry down to Z$_{2}$,
thus the effective field theory appears to be a Z$_{2}$ gauge
theory instead of U(1) in the low energy limit. As a result,
deconfinement of spinons and holons is naturally allowed in
$(2+1)D$. When spinons become gapped via further hole doping, an
anomalous metallic phase results with spin-charge separation. We
argued that there is a crossover energy scale $T_{\chi}$ from
Z$_{2}$ to U(1) above this temperature, where transport should be
more carefully studied in the presence of such gauge interactions.
We also discussed how coherent electron excitations near the
diagonal points can appear in the slave-fermion context. Lastly,
we have speculated superconductivity within the slave-fermion
representation.

\appendix*

\section{Analytic expressions for fermion self-consistent equations near zero doping}

In this appendix we show the analytic expression of $\chi_{b}$ in
the low doping limit. Minimizing the free energy Eq. (27) for
$\chi_{\psi}$, we find an equation for $\chi_{b}$ and obtain the
following expression
\bqa &&  \chi_{b} + \frac{1}{2N_L}\sum'_{k} \xi_{k}^{\psi} =
\frac{1}{8N_{L}} \Bigl[ \sum_{\xi_{k}^{\psi}\geq - (\mu +
\lambda)/2t\chi_{\psi}} \xi_{k}^{\psi} \nn && -
\sum_{\xi_{k}^{\psi}\leq - (\mu + \lambda)/2t\chi_{\psi}}
\xi_{k}^{\psi} - \sum_{\xi_{k}^{\psi} \leq (\mu +
\lambda)/2t\chi_{\psi}} \xi_{k}^{\psi} \nn && +
\sum_{\xi_{k}^{\psi} \geq (\mu + \lambda)/2t\chi_{\psi}}
\xi_{k}^{\psi} \Bigr] = \frac{1}{8N_{L}} \Bigl[
\sum_{\xi_{k}^{\psi}\geq - (\mu + \lambda)/2t\chi_{\psi}}
\xi_{k}^{\psi} \nn && + \Bigl( \sum_{(\mu + \lambda)/2t\chi_{\psi}
\leq\xi_{k}^{\psi}\leq - (\mu + \lambda)/2t\chi_{\psi}}
\xi_{k}^{\psi} + \sum_{\xi_{k}^{\psi} \geq - (\mu +
\lambda)/2t\chi_{\psi}} \xi_{k}^{\psi} \Bigr) \nn && -
\sum_{\xi_{k}^{\psi}\leq (\mu + \lambda)/2t\chi_{\psi}}
\xi_{k}^{\psi} - \Bigl( \sum_{(\mu + \lambda)/2t\chi_{\psi} \leq
\xi_{k}^{\psi} \leq - (\mu + \lambda)/2t\chi_{\psi}}
\xi_{k}^{\psi} \nn && + \sum_{\xi_{k}^{\psi} \leq (\mu +
\lambda)/2t\chi_{\psi}} \xi_{k}^{\psi}  \Bigr) \Bigr] \nn && =
\frac{1}{4N_{L}} \Bigl[ \sum_{\xi_{k}^{\psi}\geq - (\mu +
\lambda)/2t\chi_{\psi}} \xi_{k}^{\psi}  - \sum_{\xi_{k}^{\psi}\leq
(\mu + \lambda)/2t\chi_{\psi}} \xi_{k}^{\psi} \Bigr] \nn && =
\frac{1}{4N_{L}} \sum_{\xi_{k}^{\psi}\geq - (\mu +
\lambda)/2t\chi_{\psi}} \xi_{k}^{\psi} \nn && = \frac{1}{4N_{L}}
\Bigl[ \sum_{\xi_{k}^{\psi} \geq 0} \xi_{k}^{\psi} -
\sum_{\xi_{k}^{\psi}\leq - (\mu + \lambda)/2t\chi_{\psi}}
\xi_{k}^{\psi} \Bigr] \nn && = \frac{1}{2N_L}\sum'_{k}
\xi_{k}^{\psi} - \frac{1}{2} \times 4 \times
\int_{\xi_{k}^{\psi}\leq - (\mu + \lambda)/2t\chi_{\psi}}
\frac{d^2k}{(2\pi)^{2}}\xi_{k}^{\psi} \nn && =
\frac{1}{2N_L}\sum'_{k} \xi_{k}^{\psi}  - \frac{1}{2\pi^2} \Bigl[
\pi\bigl(\frac{\mu + \lambda}{2t\chi_{\psi}}\bigr)^3 -
\frac{1}{3}\pi\bigl(\frac{\mu + \lambda}{2t\chi_{\psi}}\bigr)^3
\Bigr] \nn && = \frac{1}{2N_L}\sum'_{k} \xi_{k}^{\psi} -
\frac{1}{3\pi}\Bigl(\frac{ \mu+\lambda }{2t\chi_{\psi}}\Bigr)^{3}
\eqa
with $\xi_{k}^{\psi} \equiv \sqrt{\cos^{2}k_{x} + \cos^{2}k_{y}}$.

In the same way we find the following expression from the
variation of the free energy for the chemical potential $\mu$
\bqa && - \delta = \frac{1}{4N_{L}} \Bigl[
\sum_{\xi_{k}^{\psi}\geq - (\mu + \lambda)/2t\chi_{\psi}} -
\sum_{\xi_{k}^{\psi}\leq - (\mu + \lambda)/2t\chi_{\psi}} \nn && +
\sum_{\xi_{k}^{\psi} \leq (\mu + \lambda)/2t\chi_{\psi}} -
\sum_{\xi_{k}^{\psi} \geq (\mu + \lambda)/2t\chi_{\psi}} \Bigr]
\nn && = \frac{1}{4N_{L}} \Bigl[ \sum_{\xi_{k}^{\psi}\geq - (\mu +
\lambda)/2t\chi_{\psi}} - \Bigl( \sum_{(\mu +
\lambda)/2t\chi_{\psi} \leq \xi_{k}^{\psi} \leq -
(\mu+\lambda)/2t\chi_{\psi}} \nn && + \sum_{\xi_{k}^{\psi} \geq -
(\mu + \lambda)/2t\chi_{\psi}} \Bigr) + \sum_{\xi_{k}^{\psi}\leq
(\mu + \lambda)/2t\chi_{\psi}} \nn && - \Bigl( \sum_{(\mu +
\lambda)/2t\chi_{\psi} \leq \xi_{k}^{\psi} \leq - (\mu +
\lambda)/2t\chi_{\psi}} + \sum_{\xi_{k}^{\psi} \leq (\mu +
\lambda)/2t\chi_{\psi}}\Bigr) \Bigr] \nn && = -
\frac{1}{2N_L}\sum_{(\mu + \lambda)/t\chi_{\psi}
\leq\xi_{k}^{\psi}\leq - (\mu + \lambda)/2t\chi_{\psi}} \nn && = -
\frac{1}{2N_{L}} \sum_{0 \leq \xi_{k}^{\psi} \leq - (\mu +
\lambda)/2t\chi_{\psi}} \nn && \approx - \frac{1}{2} \times 4
\times \int_{\xi_{k}^{\psi} \leq - (\mu + \lambda)/2t\chi_{\psi}}
\frac{d^2k}{(2\pi)^2} \nn && = - \frac{1}{2\pi^2}\times \pi
\bigl(\frac{\mu + \lambda}{2t\chi_{\psi}}\bigr)^2 . \eqa
As a result, we obtain \bqa && \chi_{b} = -
\frac{2\sqrt{2\pi}}{3}\delta^{3/2} . \eqa In the uniform-flux case
we find $\chi_{b} \propto - \delta^{2}$ smaller than the above in
the zero doping limit


\begin{thebibliography}{99}
\bibitem{Heavy_Fermion} In the heavy-fermion problem it is well
known that the slave-boson approach describes the heavy-fermion
Fermi-liquid phase quite well, but has difficulty in explaining
antiferromagnetism while the slave-fermion approach captures the
antiferromagnetic state well, but has difficulty in discussing the
heavy Fermi-liquid phase. This is deeply related with statistics
of spinons.
\bibitem{SO5WZW} Recently, one of the authors has derived an effective field theory
of the slave-fermion spirit from the SU(2) slave-boson theory,
where an SO(5) Wess-Zumino-Witten description for spin
fluctuations and non-relativistic fermion U(1) gauge theory for
dynamics of doped holes naturally emerge from fermionizing SU(2)
slave bosons. See Ki-Seok Kim, arXiv:0804.0895 to be published in
Phys. Rev. B.
\bibitem{S_SB_MI} It will be an interesting problem
to find a parameter discriminating symmetric Mott insulators from
symmetry-broken ones, which may be analogous to the
Ginzburg-Landau parameter distinguishing type I superconductors
from type II ones.
\bibitem{PALee_Review} P. A. Lee, Rep. Prog. Phys. {\bf 71},
012501 (2008); P. A. Lee, N. Nagaosa, and X.-G. Wen, Rev. Mod.
Phys. {\bf 78}, 17 (2006).
\bibitem{Sachdev_Review} S. Sachdev, Annals Phys. {\bf 303} 226
(2003).
\bibitem{Sachdev_BOP} K. Park and S. Sachdev, Phys. Rev. B
{\bf 64}, 184510 (2001).
\bibitem{Book} Assa Auerbach, \textit{Interacting Electrons and Quantum
magnetism }(Springer-Verlag, 1994).
\bibitem{SF_OLD} B. I. Shraiman and E. D. Siggia, Phys. Rev.
Lett. {\bf 60}, 740 (1988); P. B. Wiegmann, Phys. Rev. Lett. {\bf
60}, 2445 (1988); S. A. Trugman, Phys. Rev. B {\bf 37}, 1597
(1988); C. L. Kane, P. A. Lee, and N. Read, Phys. Rev. B {\bf 39},
6880 (1989); S. Sachdev, Phys. Rev. B {\bf 39}, 12232 (1989); X.
G. Wen, Phys. Rev. B {\bf 39}, 7223 (1989); R. Shankar, Phys. Rev.
Lett. {\bf 63}, 203 (1989); H. J. Schulz, Phys. Rev. Lett. {\bf
65}, 2462 (1990); P. A. Marchetti, G. Orso, Z. B. Su, and L. Yu,
Phys. Rev. B {\bf 71}, 134510 (2005); C. Brugger, F. Kampfer, M.
Moser, M. Pepe, and U.-J. Wiese, Phys. Rev. B {\bf 74}, 224432
(2006).
\bibitem{ACL} R. K. Kaul, A. Kolezhuk, M.
Levin, S. Sachdev, and T. Senthil, Phys. Rev. B {\bf 75}, 235122
(2007); R. K. Kaul, Y. B. Kim, S. Sachdev, and T. Senthil, Nature
Physics {\bf 4}, 28 (2008).
\bibitem{BCS_Hubbard} K.-S. Kim and M. D. Kim,
Phys. Rev. B {\bf 77}, 125103 (2008); Ki-Seok Kim and Mun Dae Kim,
Phys. Rev. B {\bf 75}, 035117 (2007).
\bibitem{NFL_SL} The spin-gapped U(1) anomalous metal in the
slave-fermion approach is quite analogous to the U(1) spin liquid
in the slave-boson representation. An effective field theory for a
doped symmetric Mott insulator is given by a non-relativistic
fermion (spinon) U(1) gauge theory for spin dynamics (of gapless
spinful fermions around the four diagonal points) and a
non-relativistic boson (holon) U(1) gauge theory for charge
dynamics, allowing the U(1) spin liquid phase (at least at finite
temperatures) where bosonic holons are gapped.
\bibitem{Deconfinement_Senthil} T. Senthil,
A. Vishwanath, L. Balents, S. Sachdev, and M. P. A. Fisher,
Science {\bf 303}, 1490 (2004); T. Senthil, L. Balents, S.
Sachdev, A. Vishwanath, and M. P.A. Fisher, Phys. Rev. B {\bf 70},
144407 (2004); M. Hermele, T. Senthil, M. P. A. Fisher, P. A. Lee,
N. Nagaosa, and X.-G. Wen, Phys. Rev. B {\bf 70}, 214437 (2004).
\bibitem{Deconfinement_Kleinert} H. Kleinert, F. S. Nogueira, and A. Sudbo, Phys.
Rev. Lett. {\bf 88}, 232001 (2002); H. Kleinert, F. S. Nogueira,
and A. Sudbo, Nucl. Phys. B {\bf 666}, 361 (2003); F. S. Nogueira
and H. Kleinert, Phys. Rev. Lett. {\bf 95}, 176406 (2005); F. S.
Nogueira and H. Kleinert, Phys. Rev. B {\bf 77}, 045107 (2008).
\bibitem{Deconfinement_Kim} Ki-Seok Kim,
Phys. Rev. B {\bf 72}, 035109 (2005); Ki-Seok Kim, Phys. Rev. B
{\bf 72}, 245106 (2005); Ki-Seok Kim, Phys. Rev. B {\bf 70},
140405(R) (2004); Ki-Seok Kim, Phys. Rev. B {\bf 72}, 014406
(2005); Ki-Seok Kim, Phys. Rev. B 72, 214401 (2005).
\bibitem{Skyrmion} It should be noted that there are two kinds of
gauge fluctuations. One are associated with "ferromagnetic" gauge
fluctuations, corresponding to phase fluctuations of a hopping
parameter in the slave-boson approach. The other are related with
"antiferromagnetic" gauge fluctuations, corresponding to phase
fluctuations of a singlet-pairing order parameter in the
Schwinger-boson approach of the Heisenberg model. Monopole
excitations in latter gauge fields correspond to skyrmions while
those in former gauge fields have nothing to do with skyrmions.
\bibitem{Polyakov} A. M. Polyakov, \textit{Gauge Fields and
Strings} (Harwood Academic Publishers, 1987).
\bibitem{Sachdev_Read} N. Read and S. Sachdev, Phys. Rev. B {\bf
42}, 4568 (1990); S. Sachdev and N. Read, Int. J. Mod. Phys. B
{\bf 5}, 219 (1991).
\bibitem{Fradkin_gauge_theory} E. Fradkin, S. H. Shenker, Phys. Rev. D {\bf 19}, 3682
(1979).
\bibitem{Order_By_Disorder} A. M. Tsvelik, Quantum Field Theory in
Condensed Matter Physics (Chap. 12, Cambridge University Press,
Cambridge, 1995).
\bibitem{AF_spinon} C.-M. Ho, V. N. Muthukumar, M. Ogata, and P.
W. Anderson, Phys. Rev. Lett. {\bf 86}, 1626 (2001), and
references therein.
\bibitem{Luttinger_Theorem} One can understand the emergence of hole pockets from the
Luttinger theorem, stating that the volume of a Fermi surface
equals to the number of fermions inside the Fermi surface.
Considering the generalized single-occupancy constraint with the
presence of spinon-singlet pairs $(\Delta_{i})$ in the
slave-fermion representation, \bqa &&
\sum_{\sigma}b_{i\sigma}^{\dagger}b_{i\sigma} +
2\Delta_{i}^{\dagger}\Delta_{i} + f_{i}^{\dagger}f_{i} = 1 ,
\nonumber \eqa we find that an electron number operator can be
written as follows \bqa && N_{el} = \sum_{\sigma}
c_{i\sigma}^{\dagger}c_{i\sigma} = \sum_{\sigma}
b_{i\sigma}^{\dagger}f_{i}f_{i}^{\dagger}b_{i\sigma} =
f_{i}f_{i}^{\dagger}(1 - f_{i}^{\dagger}f_{i} -
2\Delta_{i}^{\dagger}\Delta_{i}) \nn &&  = (1 - 2
\Delta_{i}^{\dagger}\Delta_{i})(1 - f_{i}^{\dagger}f_{i}) .
\nonumber \eqa As a result, we obtain the Luttinger theorem \bqa
&& \frac{V_{FS}^{f}}{(2\pi)^{d}} = 1 - \frac{N_{el}}{1 -
2|\Delta|^{2}} , \nonumber \eqa where $V_{FS}^{f}$ is the volume
of the holon Fermi surface and $|\Delta|^{2}$ is the condensation
amplitude of spinon pairs. When $|\Delta|^{2} << 1$ is satisfied,
one can expand the denominator, then we find in the zeroth-order
approximation for $\Delta$ \bqa && \frac{V_{FS}^{f}}{(2\pi)^{d}} =
1 - {N_{el}} = \delta .  \nonumber \eqa Thus, the volume of the
holon Fermi surface should be proportional to hole concentration,
explaining why hole pockets appear at low doping. However, the
presence of spinon-singlet pairs reduces the volume of the holon
Fermi surface, smaller than hole concentration $\delta$.
\bibitem{Spin_Value} We have used $S = 0.3$ instead of $S = 0.5$
since the present semiclassical treatment overestimates the spin
value. Quantum fluctuations will reduce such a value.
\bibitem{Don_Kim} D. H. Kim and P. A. Lee, Annals
Phys. {\bf 272}, 130 (1999).
\bibitem{ARPES} A. Damascelli, Z. Hussain, and Z.-X. Shen,
Rev. Mod. Phys. {\bf 75}, 473 (2003).
\bibitem{NL_GT} P. A. Lee and N. Nagaosa, Phys. Rev. B {\bf 46},
5621 (1992); L. B. Ioffe and G. Kotliar, Phys. Rev. B {\bf 42},
10348 (1990).
\bibitem{Nodal_electron} W. Rantner and X.-G. Wen,
arXiv:cond-mat/0105540 (unpublished).
\bibitem{SF_ALM} P. Coleman, C. Pepin, Q. Si, and R. Ramazashvili,
J. Phys. Condens. Matter {\bf 13}, R723 (2001).
\bibitem{BCS_Hubbard_SR} Ki-Seok Kim, Phys. Rev. B 75, 245105 (2007);
Ki-Seok Kim, Phys. Rev. Lett. {\bf 97}, 136402 (2006).
\end{thebibliography}
\end{document}